# On the correlation of angular distributions of keV ions and trajectory-dependent electronic excitations in transmission channelling geometry

**R Holeňák[1], S Lohmann[1,2], K Komander[1] and D Primetzhofer[1]**

[1]Department of Physics and Astronomy, Uppsala University, Box 516, S-751 20 Uppsala, Sweden
[2]Institute of Ion Beam Physics and Materials Research, Helmholtz-Zentrum Dresden-Rossendorf e.V. (HZDR), 01328 Dresden, Germany

E-mail: radek.holenak@physics.uu.se

**Abstract.** We use energy discrimination of keV ions transmitted through a thin, single-crystalline silicon membrane to correlate specific angular distribution patterns formed in channelling geometry with trajectory-dependent electronic energy loss. The integral energy and intensity distribution of transmitted ions can thus be dissected into on one side axially channelled projectiles travelling along rather straight trajectories and on the other side dechannelled projectiles predominantly experiencing blocking. Angular distributions of transmitted ions are further simulated with two different Monte-Carlo codes.

## 1. Introduction

Ever since the discovery of ion channelling [1], considerable effort has been directed towards the understanding of the specific trajectories of projectiles in the crystal lattice. In transmission geometry, ions channelled along axial directions of a single-crystal exhibit characteristic star-like angular distributions. The general appearance of these patterns and the intensity distribution within them are dependent on several interrelated parameters, such as projectile energy, mass, charge, beam divergence, but also notably on crystal thickness [2]. For MeV projectiles, the simplest patterns were observed for fairly thick targets (tens of µm) [3]. In these cases, the projectiles establish an oscillatory motion around the channel axis with characteristic maximum angle and minimum impact parameter. As the target crystal is made thinner, the majority of projectiles performs less than one oscillation and their angular distribution changes. For very thin and perfect crystals, the transmitted patterns are expected to become complex and highly dependent on specific details of the channelled trajectories inside the crystal. Several scattering theories in single crystals, pioneered by Lindhard [4], and Monte Carlo codes based on the work of Barrett [5] were used to study the projectile trajectories of MeV ions and predict the formation of patterns for transmitted projectiles [6,7]. A qualitative agreement with performed experiments [8,9] was found for some of the predicted phenomena like superfocusing [10,11] or rainbow scattering [12,13].

Ion channelling of keV ions through a thin crystal membrane is seemingly analogous to the case of MeV projectiles in a thick crystal. Multiple oscillations and inelasticity of the collisions cause significant energy loss straggling effects that hamper the observation of highly resolved transmission patterns [14]. Recently, trajectory-dependent electronic stopping power for ions with velocities around and below one Bohr velocity have been demonstrated in transmission experiments through single crystalline silicon [15,16]. We showed that projectiles experiencing a higher number of collisions with small impact parameters will suffer higher specific energy loss. These results were obtained by comparison of trajectories recorded in channelling geometry with such obtained in random orientation of the sample. In the MeV regime, the measured effects are expected to be governed by inner-shell excitations [17], which are mostly inaccessible during channelling, with limited variation for different channelling trajectories [18,19]. For keV ions heavier than hydrogen, the nature of the measured differences in energy loss is rooted in the frequent exchange of electrons between projectile and target electronic structure, including formation of molecular orbitals for heavier projectiles and the strong impact parameter dependence of these dynamic interactions [20–22]. These processes requiring close collisions cannot only explain earlier observed non-linearities in the stopping power of ions heavier than protons [23–25] but are also expected to leave a fingerprint in the energy loss along different trajectories developed during channelling.

For this reason, a more in-depth analysis of the correlation of energy loss and transmitted intensity in channelling orientations is desirable. Along with the spatial distribution, the third dimension provided by the simultaneous measurement of the projectiles flight time allows for direct association of the transmitted patterns with a specific exit energy of the detected projectiles [26]. The observed patterns in angular distributions discriminated by the energy can thus provide detailed information on the trajectory of the projectiles, the experienced impact parameter distribution and thus indirectly on the specific weight of the above stated dynamic energy loss processes itself.

Moreover, we have tested the applicability of two simulation packages, i.e. TRIC [27] and FLUX7 [28] for the prediction of angular distribution patterns obtained upon ion trasnmission in the keV energy regime. The latter has widely been used to study channelling phenomena at MeV energies.

## 2. Experiment and data processing

Experiments were performed using the medium energy ion scattering instrumentation at Uppsala University [29,30]. The particle beams are produced using a 350 kV Danfysik implanter and delivered to the chamber through a 7 m long beamline featuring a number of slits, apertures and ion optics narrowing the beam to a spot size below 1 mm x 1 mm and angular divergence of max. 0.056°. Furthermore, the electrostatic chopper shapes the beam into 1-3 ns pulses. The scattering chamber contains a large position-sensitive detector from Roentdek [31] rotatable around the central axis at a distance of 290 mm. The solid angle covers 0.13 sr corresponding to an angular width of +-11.5°. Positions of the detected particles after transmission through the sample are determined with the help of two perpendicular delay lines. The transmission foils are attached to a 6-axis goniometer necessary for accurate beam-sample alignment. The energy of the projectiles can be calculated from their measured flight times and calibrated by ion-induced photoemission [32].

Channelling patterns were studied using 150 keV $^{20}Ne^+$ primary ions transmitted through thin single-crystalline Si(100) UberFlat membranes with a nominal thickness of 50 nm purchased from Norcada Inc. [33]. The actual thickness and purity of the samples were previously assessed by Rutherford backscattering spectrometry and elastic recoil detection analysis showing a 53 nm thick high purity bulk together with a presence of light element contamination (H, C, O) on both of the surfaces [16].

Acquired data are further computationally processed. The entire area of the detector with a diameter of 120 mm is divided into square bins with a side length of 0.5 mm, which corresponds to an angular resolution of 0.1°. Each of these spatial bins contains an energy distribution histogram formed by ions arriving within the bin on the detector. Data processing allows for discrimination along this third dimension and thus a plotting sequence of 2D distribution maps of ions arriving only within a certain energy range. Moreover, the energy histograms allow for derivation of statistical information like mean energy and standard deviation, which can be assigned back to their respective positions on the detector. Details on the data processing are described in our previous work [26].

## 3. Results

Figure 1(a) depicts the experimentally recorded intensity distribution obtained from Ne ions with an initial energy of 150 keV after transmission through a 53 nm silicon crystal along the axial channel <100>. The observed distribution shows that the transmitted beam remains concentrated around the channel axis with lower intensity extending radially outward. The intensity map is plotted with a logarithmic scale to display all trajectories developed during channelling that would otherwise drown in the overwhelming intensity of projectiles detected in the direction of the beam. Only a small fraction of the projectiles escapes the channel and can be detected under larger deflection angles. Some of these dechannelled projectiles experience blocking effects, which results in a reduced intensity along the projections of the principal crystal planes at the outskirt of the detector.

The map presented in figure 1(b) is a result of the post-processing conducted on the acquired data. The mean energy is calculated for every single spatial bin and assigned to its respective position. Dependence of the projectile exit energy on the deflection angle is revealed. The energy range of all detected projectiles ranges from 120 keV almost to the initial 150 keV. Ions following trajectories along the channel or principal planes experience reduced energy loss in comparison to ions exiting the channel in random orientation.

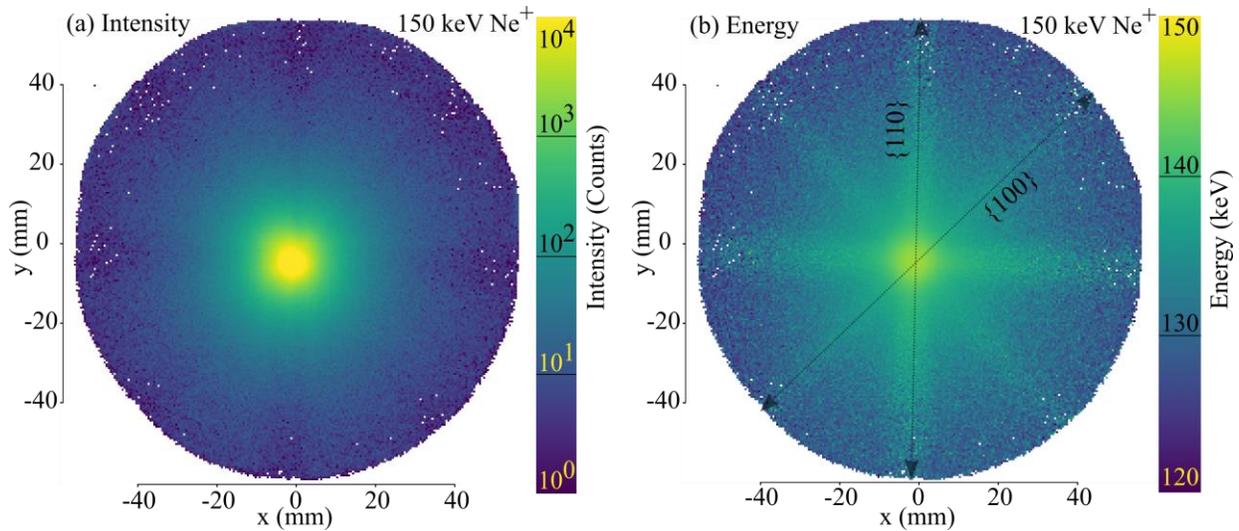

**Figure 1:** (a) Intensity and (b) Energy contrast maps for 150 keV Ne incident along the <100> channel in a 53 nm thick silicon single-crystal. Dotted arrows in the energy map mark the directions of two principal planes {100} and {110}.

For the present experiment, the energy resolution of the system was better than 2 keV, while the energy distribution of transmitted projectiles detected over the entire detector area extends over 30 keV. The shape of the energy distribution in figure 2 is asymmetrical, featuring a steep rise on the high energy side and a skewed fall with a tail towards lower energies. By slicing the energy distribution into 10 energy bins with a width of 3 keV each, the angular distribution of particles from each energy range can be visualized individually. Figure 3 shows a compilation of intensity distributions of transmitted Ne projectiles grouped according to their exit energy. The distributions in the figure are plotted in logarithmic scale normalized to the maximum intensity in each of the maps. The values in the bottom right corner disclose the fraction (in percent) of projectiles detected in the given energy range relative to the absolute number of detected projectiles.

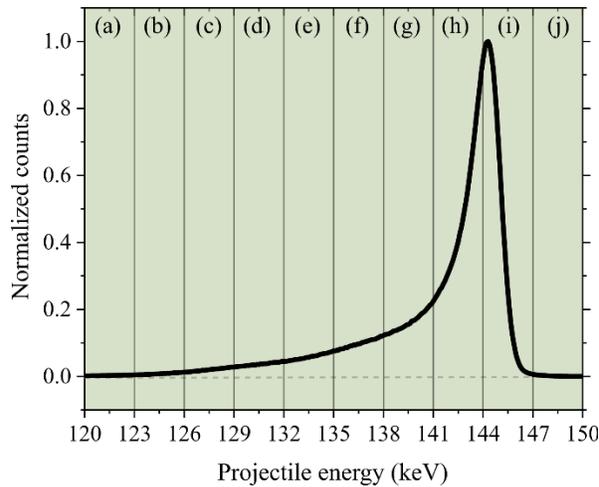

**Figure 2** Energy distribution of all detected projectiles. The full energy range of 30 keV is sliced into 10 bins with a width of 3 keV for obtaining the plots shown in figure 3.

Different structures in the intensity distribution of transmitted Ne projectiles are observed for the different energy ranges. Starting from the lowest energies, the maps which originate from the tail of the energy distribution show a wide angular spread suggesting that many of these projectiles experienced large-angle and multiple scattering. The maximum intensity is not centred on the channel axis but symmetrically extended into four quadrants separated by the low intensity streaks of the principal {110} crystal planes. With increasing energy, the maximum intensity of transmitted Ne projectiles moves towards the centre as a closing ring. Mild intensity streaks are now stretching in the direction of {110} and {100} planes resembling a star-like shape. In the energy range of 141-150 keV, the intensity distributions feature maximum intensity on the channel axis with decreasing angular spread for higher energies.

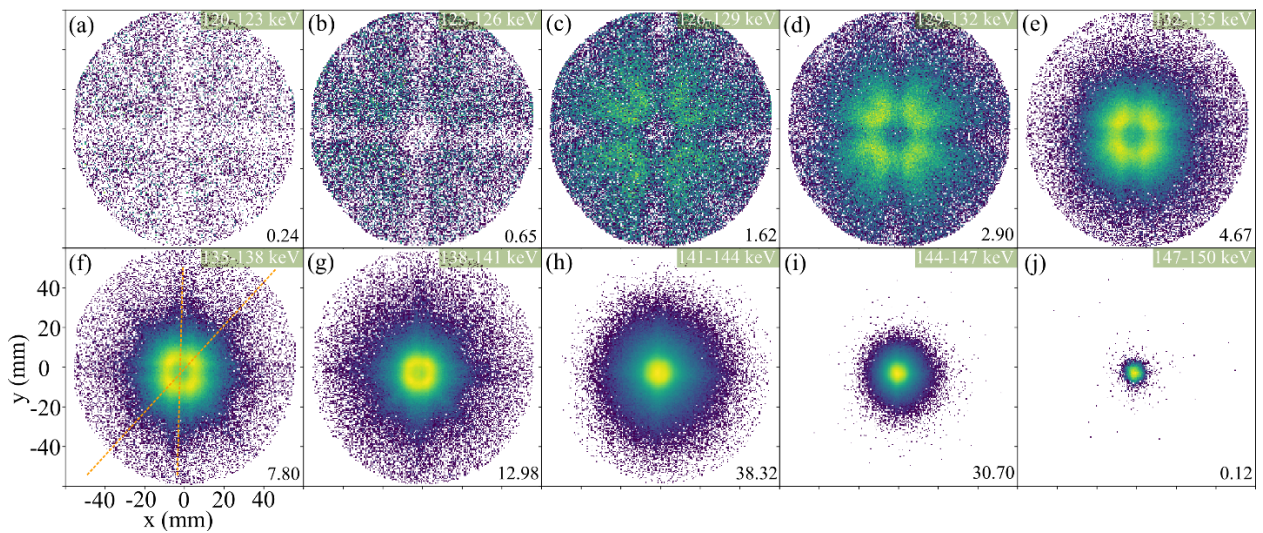

**Figure 3** Composition of scattering maps for ions arriving within given energy ranges (compare figure 2). The intensities are plotted in normalized logarithmic scales. Numbers in the right corners indicate the percentual fraction of the projectile detected in a given range relative to the absolute number of detected projectiles.

Figure 4 shows line scans taken from a selection of maps in figure 3 along the directions of two principal planes {110} and {100}. The directions of the scans are indicated by the orange dashed lines in figure 3(f). The most energetic projectiles show the narrowest angular distribution with a maximum at 0°. With decreasing energy the angular distribution broadens. Projectiles with energy below 141 keV are exiting the crystal with a non-zero scattering angle forming a ring in the observed 2D distribution and consequently a double peak in both line scans. The intensity within this ring is distributed unevenly favouring the direction of the diagonal {100} plane. All observed features lie well within the channelling half-angle of 2.48° calculated for 150 keV Ne in the silicon <100> channel [34].

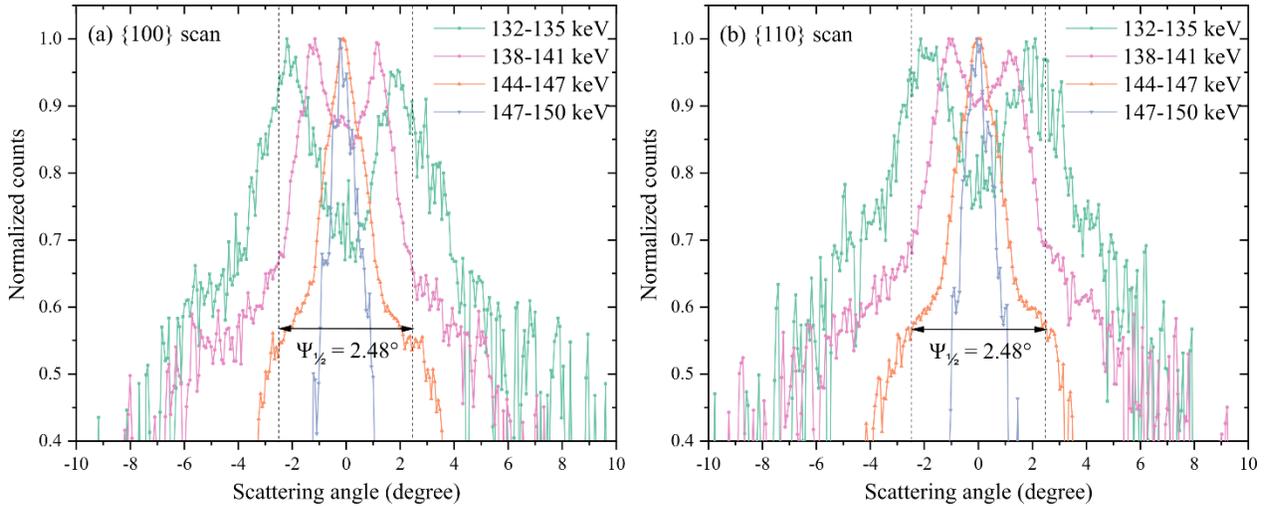

**Figure 4** Angular line scans of the channeling patterns: (a) along {100} plane, (b) along {110} plane. The dashed lines are representing the half angle calculated for 150 keV Ne.

Two Monte Carlo computational programs TRIC [27] and FLUX7 [28] were applied to simulate the ion trajectories within the Si membrane. Both codes use the Ziegler–Biersack–Littmark universal potential [35], and a binary collision model with an impact parameter dependent algorithm for energy loss. In both cases, 6 million projectiles (corresponding approximately to the number of projectiles detected in the experiment) were transmitted through a 53 nm Si crystal with the beam normal aligned to the <100> orientation. The initial particle position was at (0,0) and the beam divergence was set to 0.056°. At this stage of the work, only the overall angular distributions were studied omitting the energy dimension. Simulated intensity distributions for 150 keV Ne are presented in figure 5. In the direction of the beam, the maximum intensity is distributed in the form of a cross surrounded by fine features. Towards higher deflection angles the intensity is rapidly decreasing showing a blocking pattern. While qualitatively being very similar, for the current, not fully optimized parameters, the FLUX7 simulation predicts a wider angular spread in comparison to TRIC, similar to that observed experimentally in figure 1(a).

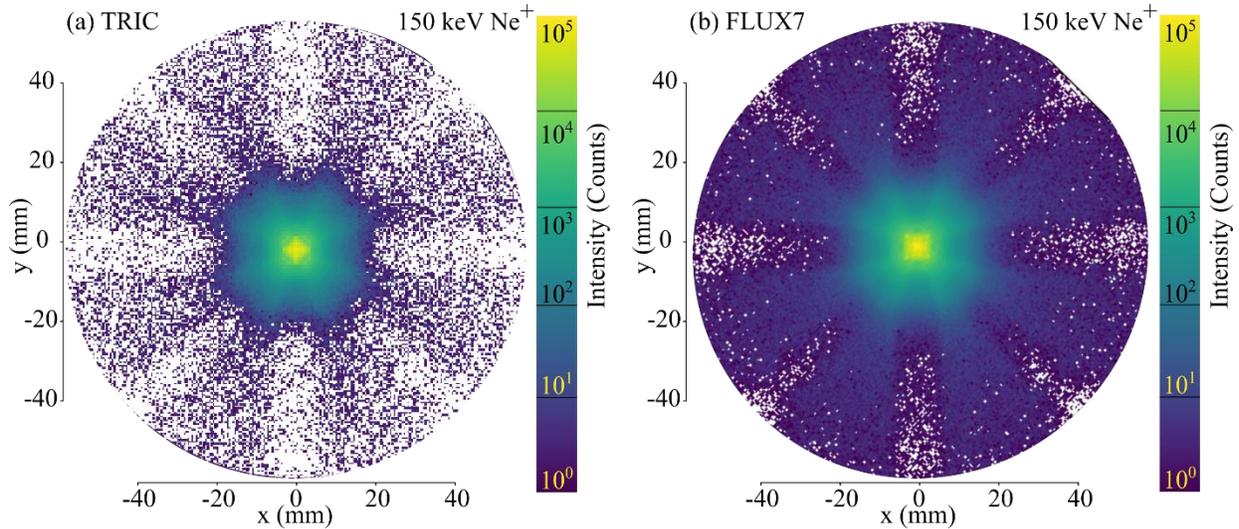

**Figure 5** Simulations of transmitted intensity patterns for 150 keV Ne and the experimental conditions by (a) TRIC code and (b) FLUX7 code.

## 4. Discussion

The patterns observed in the individual distributions provide insight into the accessed impact parameters and the specific trajectories contributing to the measured intensity: The widest angular distributions are exhibited by projectiles which upon impinging onto the sample surface do not fulfil the channelling conditions in terms of impact parameters or are dechannelled by e.g. defects. Both conditions lead to a larger deflection angle. The majority of these projectiles becomes subject to blocking that prevents them from reentering the surrounding planes. The crystalline structure as seen by the projectiles becomes analogous to random orientation. As a consequence of many close collisions, the projectiles on the way to the detector suffer higher energy loss.

Channelling occurs for projectiles entering the channel within a characteristic angular window, i.e. below a critical angle [4,34]. Any deviation from the perfect alignment with the central axis of the channel leads to the establishment of oscillatory trajectories with the oscillation period defined by the impact parameter corresponding to these angular deviations. Projectiles having the smallest acceptable impact parameter experience the highest number of oscillations slowing down the projectiles and can eventually lead to larger angular deflection on the other side of the crystal. This oscillatory motion shows a preferred orientation in the direction of {100} plane, i.e. oscillations between the two strings making up opposite corners of the channel.

Larger impact parameters allow for the development of qualitatively different channelling trajectories with longer oscillation periods leading to lower energy loss while permitting the projectile to exit into the full azimuth of the given angle. The appearing ring structure is a typically observed result of channelling in transmission experiments when the beam is purposely aligned off the channel axis but still within the critical angle for axial channelling [14,36]. In this energy range of 135-141 keV, planar channelling becomes most pronounced. Projectiles entering the planes can travel through the crystal obtaining large final deflection angles comparable to those of dechannelled projectiles, however, due to the reduced energy loss will conserve more of their initial energy. This effect is the origin of the star-like pattern in the mean energy map in figure 1(b). Finally, the most effectively channelled, i.e. least deflected ions experience the lowest energy loss and thus conserve most of their initial energy.

Good qualitative agreement is found when comparing the simulated patterns in figure 5 with the intensity distribution in figure 1(a). The differences found between the two simulation programs are expected to result from not fully optimized parameters e.g. different strengths of the screening function applied. In general, the experimental result shows wider angular scattering and the fine features in the centre of the distribution are obscured. There are multiple reasons for the quantitative disagreement such as:

- Beam divergence will effectively increase due to silicon oxide layer and surface contamination.
- TRIC simulations do not include electronic energy loss straggling.
- Thermal vibrations within a thin membrane are not necessarily the same as for bulk, in particular close to the surfaces.
- Other parameters for simulations such as importance sampling in TRIC, scattering potentials, local energy loss models, etc.

An investigation of these parameters is planned to reach a quantitative agreement.

**Summary and conclusion**
Trajectory-dependent energy loss of keV ions introduces a third dimension into the study of axially channelled projectiles in transmission geometry. The presented approach allows for a direct correlation of the energy losses with the final angular distribution. The oscillatory trajectories can reveal the details of the interaction of keV ions with crystal material. With the help of simulations, the entire trajectories and thus the number of close collisions can be extracted in the future. For the demonstration of the method we have chosen 150 keV $^{20}$Ne$^+$ due to the pronounced trajectory-dependence of the energy loss [16]. We have observed similar patterns also for other combinations of energies, ions and crystalline materials of different thicknesses.

Support of accelerator operation by the Swedish Research Council VR-RFI (Contract No. 2017-00646_9 & 2019_00191) and the Swedish Foundation for Strategic Research (Contract No. RIF14-0053) is gratefully acknowledged.